\documentstyle[preprint,aps]{revtex}
\begin{document}
\draft
\title{  Entropy in  the NUT-Kerr-Newman Black Holes Due to an Arbitrary Spin
Field }
\author{Xian-Hui Ge $^{1,2}$
You-Gen Shen$^{1,2,3}$}
\address{$^{1}$Shanghai Astronomical Observatory, Chinese Academy of Sciences,
Shanghai 200030, China (mail-address)}
\address{$^2$National Astronomical Observatories, Chinese Academy of Sciences,
Beijing 100012, China}
\address{$^3$Institute of Theoretical physics, Chinese Academy of Sciences,
Beijing 100080, China}
\date{\today}
\maketitle
 Abstract: Membrane method is used to compute the entropy of the NUT-Kerr-Newman black holes.
  It is found that even though the Euler characteristic is greater than
  two, the Bekenstein-Hawking area law is still satisfied.
  The formula $S=\chi A/8$ relating the entropy and the Euler
  characteristic becomes inapplicable for non-extreme
  four dimensional NUT-Kerr-Newman black holes.
    \hspace*{50mm}\textrm{I. Introduction}\\
 \hspace*{7.5mm}Ever since
Bekenstein and Hawking discovered that the
 entropy of a black hole is proportional to its surface area in
 the 1970s, more efforts,
  have been devoted to studying the statistical
 origin of the black hole entropy. In 1985, 't Hooft proposed brick-wall
 method and studied the statistical mechanics of a free scalar
 field in the Schwarzschild black hole background[1]. The
 reason for the introduction of the 'brick wall' method is that
 the density of states approaching the horizon diverges. Thus in
 order to avoid the divergence in the entropy, a cut-off seems has
 to be introduced, which can  be regarded as a 'brick wall'.
 Mathematically, this method requires the region of non-zero
 wave function is limited in $r_{+}+\varepsilon$ and $L$, where
 $r_{+}$ is the radius of event horizon and $\varepsilon$, $L$
 are ultraviolet cut-off and infrared cut-off respectively, and the
 relations:$\varepsilon\leq r_{+}$,$L\geq r_{+}$ should be
 satisfied. Even though it seems very arbitrary to introduce a
 cut-off in the brick wall method, it has been widely
 applied to scalar field and fermion field in various black hole
 background[2-11], where it showed that the leading term of the entropy
 for both bosons and fermions is proportional to the area of the
 event horizon.\\
\hspace*{7.5mm} Although the brick wall method achieved great
success in the study of black hole entropy,  when it is applied to
non-spherical black holes such as Kerr, and Kerr-Newman black
holes, the calculation seems rather complex[12,13]. In particular,
the brick wall method need one has to adopt small approximation
mass and remove the the term which is not proportional to area in
the integration. Moreover, if one applies the brick wall method to
Schwarzschild-de Sitter space-time[14]. Different from non-de
Sitter space-time, there are two event horizons in
Schwarzschild-de Sitter space-time. The two event horizons have
different temperatures. Therefore the radiation between them is
not in thermal equilibrium. It is clear that one should not use
the brick wall method which bases on the thermal equilibrium. In
other words, the region $[r_{+}+\varepsilon, L]$ is not in thermal
equilibrium. One should notice that former work conducted under
the brick wall method tells us that the leading term of entropy
comes from the contribution of the field very close to the
horizon. Thus one can chose a thin membrane of quantum fields in
the vicinity of each event horizon[15]. If the distance from the
membrane is $\delta$, $\varepsilon$ and $\delta$ should have the
same order. That is to say we may assume the thickness of the
membrane is also $\varepsilon$. And then the fields in the
membrane $[r_{+}+\varepsilon, r_{+}+2\varepsilon]$ can be regarded
as in locally thermal equilibrium. In fact, Hawking radiation also
comes from the vacuum fluctuation in the vicinity of event
horizon. Therefore we might regard the two event horizons as two
independent thermal equilibrium systems and consider them
respectively. This method is also applied to non-de Sitter
space-time. We  call this method 'membrane method'. The physical
picture of this method is very clear. \\\hspace*{7.5mm} On the
other hand, the membrane method enables one free from dealing with
the angular modes in the computation as if he/she can separate
field equations into two parts: the radial equations and the
angular equations. That is because just dealing with radial modes
can give the Bekenstein-Hawking area law. If the worry is about
the coefficient, then in the membrane method, as in the brick-wall
method, one has to adjust the cut-off distance to get the usual
coefficient, so missing out modes is not going to make matters
worse.\\
\hspace*{7.5mm}However, an important issue should be pointed out
here is the singularities of the metric of NUT charged spacetime,
which is called Misner strings[16]. In order to avoid the
singularities the time coordinate must be periodic. In the
Euclidean section this forces a periodicity proportional to the
NUT charge that must be matched to the usual periodicity
requirement following from the elimination of conical
singularities in the ($r,t$) section. Thus the NUT charge and the
rotation parameter must be analytically continued. Series of
papers have worked on it [17-22]. But when involving rotation in
spacetime, it is not clear that the vanishing of the metric
function at the horizon yields the same physics its non-Wick
rotated version. Fortunately, R.B.Mann managed to calculate the
entropy of Kerr-NUT class spacetime by using a boundary
counterterm prescription motivated by the AdS-CFT conjecture[23].
We will take the above issues into account in our present paper.
Moreover,the geometrical properties of several gravitational
instantons have been investigated in [24]. As it was speculated by
some authors that the geometrical properties play an essential
role in the explanation of intrinsic thermodynamics of black holes
and entropy of a black hole can be expressed in the following
formula: $S=A\chi/8$[24], in this paper, we will go further to
study the entropy of the NUT-Kerr-Newman black holes. Their
entropy of spin fields is calculated by using the membrane method
( The gravitational field doesn't be taken into account because it
seems difficult to separate the field equations). The results show
that as the cut-off is properly chosen, the entropy in the black
 hole satisfies the Bekenstein-Hawking area law.  What is more, it seems that the formula $S=A\chi/8$ has
its limitations: the author only  discusses its application to
four dimensional sphere-topology black holes.\\
 \hspace*{7.5mm} Our paper is organized as follows. In the next section, we
 discuss the singularities of
 the NUT-Kerr-Newman metric and make some mathematic provision. We
 sperate equations in SecIII. Then we will go ahead, in SecIV,
 with the reduction of the radial part to the one-dimensional wave
 equation through a series of transformations, we calculate the
 entropy. In the last section, we present our conclusions.\\
 \hspace*{50mm}\textrm{II. NUT-Kerr-Newman metric}\\
  \ \hspace*{7.5mm}The NUT-Kerr-Newman space-times can be written
in Boyer-Lindquist coordinates as [25](here $\Lambda=0$)
 \begin{eqnarray}
 &&ds^{2} = \frac{{\Delta  -  a^{2}sin^{2}\theta} }{{\rho
^{2}}}dt^{2} + 2\frac{{ \left( {r^{2} + a^{2}}
\right)asin^{2}\theta - \Delta\left( {a - \frac{{\left( {n -
acos\theta} \right)^{2}}}{{a}}} \right)}}{{\rho ^{2}}}dtd\phi
\nonumber\\&&- \frac{{\rho ^{2}}}{{\Delta } }dr^{2} - {{\rho
^{2}}}d\theta ^{2} \
 - \frac{{\left( {r^{2} + a^{2}} \right)^{2}sin^{2}\theta
- \Delta\left( {a - \frac{{\left( {n - acos\theta}
\right)^{2}}}{{a}}} \right)^{2}}}{{\rho ^{2}}}d\phi ^{2},
\end{eqnarray}
where $\rho^{2}, \bar {\rho},\Delta$, are defined by
\begin{eqnarray}
\rho ^{2} &=& \bar {\rho}  \cdot \bar {\rho}  * ,   \bar {\rho}  =
r +
i\left( {acos\theta - n} \right), \nonumber \\
 \Delta &=&  {r^{2} + a^{2} - n^{2}} +Q^2- 2Mr ,
 \end{eqnarray}
where $M,a,$ and $n$,are the mass, angular momentum per unit mass,
and the NUT parameter (n is also called gravitational magnetic
type mass). This metric has a singularity at ${\theta=0}$ and
$\theta=\pi$. The $\cos\theta$ term in the metric means that a
small loop around the axis does not shrink to zero length at
$\theta=0$ and at $\theta=\pi$. This singularity can be regarded
as the analogue of a Dirac string in electrodynamics, and is
called the Misner string.\\ \hspace*{7.5mm}The Eucledian
NUT-Kerr-Newman instanton is
\begin{eqnarray} &&ds^{2} = \frac{{\Delta  +  a^{2}sin^{2}\theta}
}{{\rho ^{2}}}dt^{2} + 2\frac{{ \left( {r^{2} - a^{2}}
\right)asin^{2}\theta - \Delta\left( {a -\frac{{\left( {n +
acos\theta} \right)^{2}}}{{a}}} \right)}}{{\rho ^{2}}}dtd\phi
\nonumber\\&&+ \frac{{\rho ^{2}}}{{\Delta } }dr^{2} +{{\rho
^{2}}}d\theta ^{2} \
 +\frac{{\left( {r^{2} - a^{2}} \right)^{2}sin^{2}\theta
+ \Delta\left( {a - \frac{{\left( {n + acos\theta}
\right)^{2}}}{{a}}} \right)^{2}}}{{\rho ^{2}}}d\phi
^{2},\nonumber\end{eqnarray} where \begin{eqnarray} \rho ^{2}&=&
r^{2}-(n+a\cos\theta)^{2},\nonumber \\
\Delta &=&  {r^{2} + a^{2} - n^{2}} +Q^2-
2Mr.\nonumber\end{eqnarray} \hspace*{7.5mm}It was argued by
Gibbons and Hawking that there are two kinds of basic
gravitational instantons based on the dimension of the fixed point
set of the continuous isometry group[26]. The first kind contains
an isolated fixed point. The second type contains a 2-dimensional
fixed point set is called a bolt. It also has been shown that
entropy can be associated with a broader and qualitatively
different gravitational system, which contains Misner strings[20].
Gravitional entropy arises whenever it is not possible to foliate
a given spacetime in the Euclidean regime by a family of surfaces
of constant time $\tau$. Such break down in foliation can occur if
the U(1) has a fixed point set of even co-dimension $d_{f}<d-2$
(called a nut)[26]. If the fixed point set has co-dimension d-2
then the usual relationship between entropy and area holds.
However, if there exists additional fixed point sets with lower
co-dimensionality then the relationship between area and entropy
is generalized. Following the idea of [23],we find the location of
the nut is at $r=a=r_{n}$ and the surface gravity is given by
\begin{eqnarray}\kappa=\frac{(r_{+}-r_{-})}{2(r^{2}_{+}-a^{2})}\nonumber
\end{eqnarray}
Here $r_{+}$ and $r_{-}$ satisfies $\Delta=(r-r_{+})(r-r_{-})=0$.
The regularity in the (r,$\tau$) section implies that $\tau$ has
period $2\pi/\kappa$. If we proceeding along this line, we can
obtain the entropy of scalar field in Euclidean NUT-Kerr-Newman
background that it is $\pi(r^{2}_{+}-a^2)$, which is in agreement
with Ref.[23].
\\\hspace*{7.5mm} Here, we turn to calculate the entropy of
NUT-Kerr-Newman black holes by the improved brick wall method in
curvature spacetime. The null-vectors of the Newman-Penrose
formalism [27]
 we take
 \begin{eqnarray}
 l^{\mu}&  =& \left[ {\frac{{\left( {r^{2} + a^{2}} \right)}}{{\Delta
}},1,0,\frac{{a}}{{\Delta } }} \right] ,\nonumber\\
 n^{\mu}  &= &\frac{{1}}{{2\rho ^{2}}}\left[ {\left( {r^{2} + a^{2}} \right),
- \Delta  ,0,a} \right],\nonumber \\
 m^{\mu}  &=& \frac{{1}}{{\sqrt {2 }  \bar {\rho} }}\left[
{\frac{{i\left( {a - \frac{{\left( {n - acos\theta}
\right)^{2}}}{{a}}} \right)}}{{sin\theta} },0,1
,\frac{{i}}{{sin\theta} }} \right].
\end{eqnarray}
\hspace*{7.5mm}We find that the non-vanishing spin-coeffients [27]
are:
 \begin{eqnarray}
 \pi& =& \frac{{ia }  sin\theta }{{\sqrt {2} \rho
^{2}}};\mu = - \frac{{\Delta} }{{2\rho ^{2}\bar {\rho} ^{ *}
}},\alpha = \pi - \beta ^{ *} ;\beta = \frac{{cos\theta}}{{2\sqrt
{2} \bar {\rho} sin\theta }};\nonumber\\\tau &=& - \frac{{ia}
sin\theta }{{\sqrt {2} \rho ^{2}}};\rho = - \frac{{1}}{{\bar
{\rho} ^{ *} }};
 \gamma = \frac{{1}}{{4\rho ^{2}}}\frac{{d\Delta} }{{dr}} + \mu .
 \end{eqnarray}
 \hspace*{7.5mm}Assuming that the azimuthal and time dependence of our
  fields will be of the form $e^{i(m\phi-\omega{t})}$, we find that the directional derivatives
  are
\begin{eqnarray}
D=l^{\mu}\partial_{\mu}=D_{0},\Delta=n^{\mu}\partial_{\mu}=\frac{-\Delta}{2\rho^{2}}D_{0}^{+},\nonumber\\
\delta=m^{\mu}\partial_{\mu}=\frac{1}{\bar{\rho}\sqrt{2}}L_{0}^{+},
\delta^{*}=m^{*\mu}\partial_{\mu}=\frac{1}{\bar{\rho}^{*}\sqrt{2}}L_{0}
\end{eqnarray}
where $D_{n}, D_{n} ^{ +}, L_{n},  L_{n} ^{ +} , K,H $ are defined
by
\begin{eqnarray}
D_{n} &=& \partial _{r} + \frac{{i K}}{{\Delta} } +
\frac{{n}}{{\Delta} }\frac{{d\Delta} }{{dr}},\nonumber \\
 D_{n} ^{ +} &=& \partial _{r} -\frac{{i K}}{{\Delta } } +
\frac{{n}}{{\Delta } }\frac{{d\Delta } }{{dr}},\nonumber \\
 L_{n} &=& \partial _{\theta}  + H +
n{cot\theta},\nonumber \\
 L_{n} ^{ +}  &=& \partial _{\theta}  - H+n{cot\theta},\nonumber\\
 K &=&  am-\omega\left( {r^{2} + a^{2}} \right) ,\nonumber\\
 H &=& \frac{{m}}{{sin\theta}
 }-\frac{a^{2}-\left({n-acos\theta}\right)^{2}}{asin\theta}\omega
 .
 \end{eqnarray}Thus K and H have the relation
 \begin{eqnarray}
 K-aHsin{\theta}=-\rho^{2}\omega.
 \end{eqnarray}
 These differential operators satisfy some identities
 \begin{eqnarray}\Delta {D_{n+1}}&=&D_{n}\Delta\\
\Delta {D_{n+1}^{+}}&=&D_{n}^{+}\Delta.\\
 \left(sin{\theta}\right)L_{n+1}&=&L_{n}sin{\theta}\\
\left({sin{\theta}}\right)L_{n+1}^{+}&=&
L_{n}^{+}sin{\theta}\\
\left(D+\frac{m}{\bar\rho^{*}}\right)\left(L+\frac{i m a
sin{\theta}}{\bar\rho^{*}}\right)&=& \left(L+\frac{im a
sin{\theta}
}{\bar\rho^{*}}\right)\left(D+\frac{m}{\bar\rho^{*}}\right)
 \end{eqnarray}
  \hspace*{10mm}\textrm{III. Spin fields  in NUT-Kerr-Newman space-time}\\
 \hspace*{7.5mm}The Maxwell equations in the Newman-Penrose
 formalism take on the forms[27]
 \begin{eqnarray}
 D\phi_{1}-\delta^{*}\phi_{0}&=&(\pi-2\alpha)\phi_{0}+2\tilde{\rho}\phi_{1}-\kappa\phi_{2},\\
 D\phi_{2}-\delta^{*}\phi_{1}&=&-\lambda\phi_{0}+2\pi\phi_{1}+(\tilde{\rho}-2\epsilon)\phi_{2},\\
 \delta\phi_{1}-\Delta\phi_{0}&=&(\mu-2\gamma)\phi_{0}+2\tau\phi_{1}-\sigma\phi_{2},\\
\delta\phi_{1}-\Delta\phi_{0}&=&-\nu\phi_{0}+2\mu\phi_{1}+(\tau-2\beta)\phi_{2}.
 \end{eqnarray}
 Using Eqs.(4)and (5),then making the transformations
 \begin{eqnarray}
 \phi_{0}=\Phi_{0},\phi_{1}=\frac{1}{\sqrt{2}\bar\rho^{*}}\Phi_{1}\nonumber\end{eqnarray}
 and\begin{eqnarray} \phi_{2}=\frac{1}{(2{\bar\rho^{*}})^{2}}\Phi_{1},\nonumber
 \end{eqnarray}
we find that Eqs.(13)-(16) become
\begin{eqnarray}
\left[D_{0}+\frac{1}{\bar\rho^{*}}\right]\Phi_{1}&=&\left[L_{1}-\frac{i
asin{\theta} }{\bar\rho^{*}}\right]\Phi_{0}, \\
\left[D_{0}-\frac{1}{\bar\rho^{*}}\right]\Phi_{2}&=&\left[L_{0}+\frac{i
asin{\theta} }{\bar\rho^{*}}\right]\Phi_{1}, \\
\left[L_{0}^{+}+\frac{i asin{\theta}
}{\bar\rho^{*}}\right]\Phi_{1}&=&-\Delta\left[D_{1}^{+}-\frac{1}{\bar\rho^{*}}\right]\Phi_{0},\\
\left[L_{1}^{+}-\frac{i asin{\theta}
}{\bar\rho^{*}}\right]\Phi_{2}&=&-\Delta\left[D_{0}^{+}+\frac{1}{\bar\rho^{*}}\right]\Phi_{1}.
\end{eqnarray}
From Eqs.(18)and(20),$\Phi_{1}$ can be eliminated to give
\begin{eqnarray}
\left(\Delta
D_{1}D_{1}^{+}+L_{0}^{+}L_{1}-2i\omega\bar\rho\right)\Phi_{0}=0
\end{eqnarray}
Similarly, from Eqs.(19)and(21)there is
\begin{eqnarray}
\left(\Delta
D_{0}^{+}D_{0}+L_{0}L_{1}^{+}+2i\omega\bar\rho\right)\Phi_{2} =0.
\end{eqnarray}Assuming $\Phi_{0}=R_{+1}(r)S_{+1}(\theta)$ and $\Phi_{2}=R_{-1}(\theta)S_{-1}(\theta)$
, we can separate the variables of Eqs.(22) and (23) to be
\begin{eqnarray}
\left(\Delta {D_{1}D_{1}^{+}}-2ir{\omega}\right)R_{+1}={\lambda}R_{+1},\\
\left(\Delta{D_{0}^{+}D_{0}}+2ir{\omega}\right)R_{-1}={\lambda}R_{-1},\\
\left[L_{0}^{+}L_{1}+2\omega(acos\theta-n)\right]S_{+1}=-{\lambda}S_{+1},\\
\left[L_{0}L_{1}^{+}-2\omega(acos\theta-n)\right]S_{-1}=-{\lambda}S_{-1},
\end{eqnarray}
here$\lambda$ is the separation constant. For the Dirac field, the
wave equations for a massless dirac particles are[28]
\begin{eqnarray}
 \left( {D + \varepsilon - \rho}  \right)F_{1} + \left( {\bar {\delta}  +
\pi - \alpha}  \right)F{}_{2} = 0,\nonumber \\
 \left( {\Delta ^{'} - \mu - \gamma}  \right)F_{2} + \left( {\delta + \beta
- \tau}  \right)F_{1} = 0,\nonumber \\
 \left( {D + \varepsilon ^{ *}  - \rho ^{ *} } \right)G_{2} - \left( {\delta
+ \pi ^{ *}  - \alpha ^{ *} } \right)G_{1} = 0 ,\nonumber\\
 \left( {\Delta ^{'} - \mu ^{ *}  - \gamma ^{ *} } \right)G_{1} - \left(
{\bar {\delta}  + \beta ^{ *}  - \tau ^{ *} } \right)G_{2} = 0,
\end{eqnarray}
where $F_{1},F_{2},G_{1},G_{2}$ are 4-component
spinors,$\alpha,\beta,
 \gamma,\epsilon,\mu,\pi,\rho,\tau$ etc are Newman-Penrose symbols,and $\alpha^{*},\beta^{*}$
 are the complex conjugates of $\alpha,\beta$ etc.\\
\hspace*{7.5mm}All the above equations are also separated by using
Newman-Penrose formalism.The radial equations are given by
\begin{eqnarray}
 \Delta ^{\frac{{1}}{{2}}}D_{0} ^{ +} \left(
{\Delta  ^{\frac{{1}}{{2}}}D_{0} R_{ - \frac{{1}}{{2}}}} \right)
&=& \lambda ^{2}R_{ -
\frac{{1}}{{2}}} , \\
\Delta ^{\frac{{1}}{{2}}}D_{0} \left( {\Delta
^{\frac{{1}}{{2}}}D_{0} ^{ +} R_{\frac{{1}}{{2}}}}  \right) &=&
\lambda
^{2}R_{ + \frac{{1}}{{2}}} ,\\
 L_{\frac{{1}}{{2}}} ^{ +} \left(
L_{\frac{{1}}{{2}} S_{ +
\frac{{1}}{{2}}}}  \right) &=& - \lambda ^{2}S_{ + \frac{{1}}{{2}}} , \\
 L_{\frac{{1}}{{2}}} \left( L_{\frac{{1}}{{2}}} ^{ +} S_{ -
\frac{{1}}{{2}}}  \right) &=& - \lambda ^{2}S_{ -
\frac{{1}}{{2}}},
\end{eqnarray}
\hspace*{7.5mm}For scalar field, the separated equations can
achieved directly from the Klein-Gordon equation
\begin{equation}
\frac{{1}}{{\sqrt { - g}} }\frac{{\partial} }{{\partial x^{\mu}
}}\left( {\sqrt { - g} g^{\mu \nu} \frac{{\partial \Phi}
}{{\partial x^{\nu} }}} \right) = 0.
\end{equation}
The radial equation is
\begin{equation}
\frac{\partial}{\partial{r}}\Delta\frac{\partial}{\partial{r}}R(r)+\frac{\left[\left(r^2+a^2\right)
\omega-am\right]^2}{\Delta}R(r)=\lambda^{2}R(r).
\end{equation}We can rewrite the above equation with $D_{n},
D_{n}^{+}  $
\begin{equation} \Delta{D_{0}D_{0}^{+}}R_{0}=\lambda^{2}R_{0}.
\end{equation}
\hspace*{7.5mm}The radial equations (23),(29)and (34) can be
combined into
\begin{equation}
\left[\Delta D_{s}D_{s(2s-1)}^{+}-2s(2s-1)i\omega
r\right]R_{s}=\lambda^{2}R_{s}.
\end{equation}
On the other hand, equations (24) and (28) can be written as
\begin{equation} \left[\Delta D_{1-s}^{+}D_{0}+2s(2s-1)i\omega
r\right]R_{-s}=\lambda^{2}R_{-s}.
\end{equation}where
$s$ is the spin number. It is clear that as $s=0,s=\frac{1}{2}$
and $s=1$,Eq.(35) corresponds to scalar,Dirac and
Maxwell field respectively.\\
\hspace*{60mm}\textrm{IV. Entropy}\\
 \hspace*{7.5mm}By using the
Wentzel-Kramers-Brillouin approximation and substituting equations
$R_{s}=e^{if(r)}$ into Eqs.(35), the wave numbers are obtained as
follows,\begin{equation}
k_{r}^{2}=\frac{K^{2}}{\Delta^{2}}+\frac{2s(2s-1)}{\Delta}+\frac{s(2s-1)(s-1)}{\Delta^{2}}
{\left(\frac{d\Delta}{dr}\right)}^{2}-\frac{\lambda^{2}}{\Delta}.
\end{equation}Considering  while $s=0,\frac{1}{2},1$ the third term of
the equation is zero, we remove it from our equation. Therefore we
have
\begin{equation}
k_{r}=\frac{1}{\Delta}\sqrt{[(r^{2}+a^{2})\omega-am]^{2}+2\Delta
s(2s-1) -\Delta \lambda^{2}}.
\end{equation}
The horizon equation can be written as
\begin{eqnarray}
 \Delta &=& \left( {r
- r_{ +} } \right)\left( {r - r_{ -} } \right)=0,
\end{eqnarray}
where $r_{ +},r_{ -}, $ are the radius of  the black hole event
horizon, the inner horizon respectively. We assume that
 the boson field is in the Hartle-Hawking vacuum state. The spectrum of Hawking radiation in NUT-Kerr-Newman space-time
 can be written as \begin{equation}
 N_{\omega}^2=\frac{1}{e^{\beta_{+}(\omega-m\Omega_{+})}\pm1},\end{equation}where $N_{\omega}$,$\beta_{+}$,$\omega$
 and $\Omega_{+}$ are the radiation intensity, the inverse of Hawking temperature,
 the energy of particles, and the angular velocity of event
 horizon. Moreover,
 the
 angular velocity,
 the Hawking temperature of the black hole event horizon are defined by
 \begin{eqnarray}
 \Omega_{+}=\lim_{r\rightarrow r_{H}}\frac{-g_{t\phi}}{g_{\phi\phi}}  =  \frac{{a}}{{r_{+}^{2} + a^{2}}},\\
  T_{ +} =\frac{\kappa}{2\pi}= \frac{{1}}{{\beta _{ +} } } =  \frac{{\left( {r_{ +}  - r_{ -} }  \right)}}{{4\pi \left( {r_{ +}  ^{2} + a^{2}}
\right)}},
\end{eqnarray}The $\beta _{ +}$ here is in agreement with that of
[25]considering $\Lambda=0$. \\
\hspace*{7.5mm}The distribution of particles corresponding to
Eq.(40) is given by
\begin{equation}a_{l}=\frac{\omega_{l}}{e^{\beta(\omega-\omega_{0})\pm1}},\end{equation}where $a_{l}$
is the number of particles in the $l$ energy level, $\omega_{l}$
is the degeneracy of $l$ energy level; $\beta$ and $\omega_{0}$
denote $\beta_{+}$ and $\Omega_{+}$. It is clear that Eq.(43) will
be meaningless if $\omega<\omega_{0}$, because $a_{l}$ will be
negative for bosons in such case. Thus this demands $\omega$ in
Eq.(40) satisfies $\omega-m\Omega_{+}>0$ for boson fields.
 However, just as it
was pointed out by R. B. Mann in Ref.[23] that the Euclidean time
variable should be taken to be periodic, so the Lorentzian time
variable be periodic with period $8\pi$n. Therefore solutions to
the wave equation must also be periodic, and so $\omega$ in
Eq.(33) must be $C/(4n)$, where $C$ is an integer. And also this
demands in the following calculation the sum of $\omega$ should be
replaced by the sum of $C$. Considering Eq.(40) we set
$E=C/(4n)-m\Omega_{+}$. Further calculation will show that this
does not change our results if we take $E$ as  the system energy.
Thus the wave numbers refer to the horizon can be written as
\begin{equation}
 k_{r}=
\frac{{1}}{{\Delta } }\sqrt {\left( {r^{2} + a^{2}}
\right)^{2}\left( {E + m\Omega _{ +}  - m\Omega}
\right)^{2}+2\Delta{s(s-1)} - \Delta  \lambda^{2}},
 \end{equation}
 where $ \Omega = \frac{a}{r^{2}+a^{2}}$.The free energy at temperature $T_{+}$ of the boson system is
given by\begin{equation}
 \beta _{ +}  f = -
\sum\limits_{E} {ln\left( {1 \pm e^{ - \beta _{ +}  E}} \right)},
\end{equation}where $+$ corresponds to fermion field and $-$ corresponds to boson field.\\
\hspace*{7.5mm}According to semi-classical quantum theory, there
is\begin{eqnarray}
 \sum\limits_{E} {} \buildrel {} \over \longrightarrow
\int\limits_{0}^{\infty}  {dEg\left( {E} \right)} ,\nonumber
\end{eqnarray} where $g\left( {E} \right) = \omega'\frac{{d\Gamma \left(
{E} \right)}}{{dE}}$ is the states density. $\omega'$ is the
degeneracy of the fields(for scalar field and neutrino field,
$\omega'=1$; for Maxwell field, $\omega'=2$). The states number is
\begin{eqnarray} \Gamma \left( {E} \right) &=& \sum\limits_{m,\lambda}
{n_{r}} \left( {E,\lambda,m} \right)\ = \int {dm\int {d\lambda}}
\frac{{1}}{{\pi} }\int { k_{r} \left( {E,\lambda} \right)} dr.
\end{eqnarray}The free energy can be
calculated as follows\begin{eqnarray}-\beta _{ +}  f &=& \pm
\int\limits_{0}^{\infty}  {dEg\left( {E}
\right)ln\left( {1 \pm e^{ - \beta _{ +}  E}} \right)},\nonumber\\
&=& \pm \beta _{ +} \int\limits_{0}^{\infty}
{dE\omega'\frac{{\Gamma
\left( {E} \right)}}{{e^{\beta _{ +}  E} \pm 1}}}\nonumber \\
 &=&  \frac{{\beta _{ +} }\omega '}{{\pi} }\int\limits_{0}^{\infty}
{dE\int\limits_{r_{ +}  + \varepsilon} ^{r_{ +}  + 2\varepsilon}
{dr}} \int\limits_{0}^{\lambda_{max}}  {d\lambda} \int\limits_{ -
\lambda}^{\lambda} {d m}
 \frac{{1}}{{\Delta } }\left( {e^{\beta _{ +}
E} \pm 1} \right)^{ - 1}\nonumber\\ & & \sqrt {\left( {r^{2} +
a^{2}} \right)^{2}\left( {E + m\Omega _{ +}  - m\Omega}
\right)^{2}+2\Delta{s(s-1)} - \Delta \lambda^{2}},
\end{eqnarray} where the separation constant $\lambda$ is the angular quantum
 number which corresponds to $l$ in the spherical space-time case. In this case, the range of magnetic quantum
  number $m$ is $-\lambda\leq m \leq\lambda$ . $\lambda_{max}$ corresponds to the fact that $k_{r}\geq 0$ (while
  $k_{r}=0$, $\lambda$ reach its maximum).
   Considering fermions field and bosons field, the
results can be written as 
\begin{eqnarray}
f_{f}&=&- \frac{{7}}{{180}} \cdot \frac{{\pi ^{3}}}{{\beta ^{4}}}
\cdot \frac{{\left( {r_{+}^{2} + a^{2}}
\right)^{3}\omega'}}{{\left( {r_{ +}  - r_{ -} } \right)^{2}} }
\cdot \frac{{\varepsilon} }{{\eta
^{2}}}, (fermions field)\\
 f_{b}&=&- \frac{{2}}{{45}} \cdot \frac{{\pi ^{3}}}{{\beta ^{4}}} \cdot
\frac{{\left( {r_{+}^{2} + a^{2}} \right)^{3}}\omega'}{{\left(
{r_{ +}  - r_{ -} } \right)^{2}}} \cdot \frac{{\varepsilon}
}{{\eta ^{2}}},  (bosons field)
\end{eqnarray} where $\varepsilon$ is the ultraviolet regulator,
which satisfies $ 0<\varepsilon\ll r_{+}$. This manifests  that
the integral over the quantum number $m$ does not diverge,
therefore we need not to regularize the $m$ integral. On the other
hand, the membrane model illustrates that the black hole entropy
mainly comes from the vicinity of event horizon. Thus we have
taken into account the following equation in the integration with
respect to $m$,\begin{equation}\mathop {lim}\limits_{r \to r_{+} }
\Omega =\Omega_{+}.\end{equation}We also used the median theorem
in the integration with respect to $r$, hence
$\varepsilon<\eta<2\varepsilon$.
Eq.(50) is also used in the integration with respect to $r$.\\
\hspace*{7.5mm}We are now ready to obtain the entropy  due to
arbitrary spin field of the NUT-Kerr-Newman black hole from the
standard formula\begin{equation}
 S = \beta ^{2}\frac{{\partial F}}{{\partial \beta} }.
 \end{equation}
  As to fermion field, one componential entropy can be written as
   \begin{equation}S_{1f} =
\frac{{7}}{{45}} \cdot \frac{{\pi ^{3}}}{{\beta ^{3}}} \cdot
\frac{{\left( {r_{+}^{2} + a^{2}} \right)^{3}\omega'}}{\left( {r_{
+}  - r_{ -} } \right)^{2}} \cdot \frac{{\varepsilon} }{{\eta
^{2}}},\end{equation} There are four components of the wave
function refer to fermion fields. Therefore the whole black hole
entropy is given by\begin{equation}S_{ f}  = 4S_{1f} =
\frac{{28}}{{45}} \cdot \frac{{\pi ^{3}}}{{\beta ^{3}}} \cdot
\frac{{\left( {r_{+}^{2} + a^{2}} \right)^{3}\omega'}}{{\left(
{r_{ +}  - r_{ -} } \right)^{2}}} \cdot \frac{{\varepsilon}
}{{\eta ^{2}}}.
 \end{equation}
 Similarly, the entropy of boson fields can be obtained as
  \begin{equation}S_{b} =
\frac{{8}}{{45}} \cdot \frac{{\pi ^{3}}}{{\beta ^{3}}} \cdot
\frac{{\left( {r_{+}^{2} + a^{2}} \right)^{3}\omega'}}{{\left(
{r_{ +}  - r_{ -} }  \right)^{2}}} \cdot \frac{{\varepsilon}
}{{\eta ^{2}}},\end{equation} Eq.(51) is in same form as that in
[18] We choose the cut-off as $\frac{{1}}{{\varepsilon} } =
90\beta
 $. Here $\varepsilon$ and $\eta$ in Eq.(47) and Eq.(48) are of the same
 order. Therefore $\frac{\varepsilon}{\eta^{2}}\sim
\frac{1}{\varepsilon}= 90\beta
 $, then the entropy in Eq.(47) and Eq.(48) satisfies the area law\begin{eqnarray}
 S_{ f} &=& \frac{{7}}{{8}} \cdot 4\pi \left( {r_{ +}  ^{2} + a^{2}}
\right)\omega' = \frac{{7}}{{8}}A_{ +}\omega'  ,\\
 S_{ b} & =& \frac{{1}}{{4}} \cdot 4\pi \left( {r_{ +}  ^{2} + a^{2}}
\right)\omega' = \frac{{1}}{{4}}A_{ +}\omega'  ,
\end{eqnarray} where $A_{ +}$ is the area
of black hole event horizon. The results in Eq.(55) is in
agreement with our results in Ref.[29] considering $\Lambda=0$.
\\
\hspace*{60mm}\textrm{V. Discussion and Conclusion}\\
\hspace*{7.5mm}We have discussed the issue that arises
singularities in NUT-Kerr-Newman spacetime and have studied the
entropy due to arbitrary spin field in the NUT-Kerr black holes
whose Euler's characteristic is over two. Our results is in
agreement with the results in [23]. Since the cut-off was properly
chosen, the NUT-Kerr-Newman black hole entropy is identified with
the Bekenstein-Hawking area law. As the topology of the
NUT-Kerr-Newman black hole is special, and its Euler's
characteristic is greater than two, we can see that the formula $
S = \frac{{1}}{{8}}\chi A$ is not applicable to our case.
Therefore, it means that this formula in [24] has its limitations:
the equation does not apply to high dimensional black holes ($\chi
\left( {s^{2\kappa + 1}} \right) = 0$), even does not apply to
four dimensional  black holes. The NUT-Kerr-Newman black hole is
such an example.Therefore the relations between a black hole
topology and its entropy need further investigation. In addition,
we can see from the results that the electromagnetic, Dirac and
scalar field entropies of the following black holes: Schwarzschild
black hole, Reissner-Nordsrt\"{o}m black hole, Kerr black hole and
NUT-Kerr black hole are embodied as special cases of the
NUT-Kerr-Newman
black hole entropy.  \\
{\large\bf Acknowledgments:} The authors (X.H.Ge) would like to
thank the anonymous referees for their insightful comments. This
work has been supported by the National Natural Science Foundation
of China under Grant No. 10273017 and No.10073006, and the
Foundation of Shanghai Development for Science and Technology
under Grant No.01-JC14035.
\\
\\

\end{document}